\documentstyle[cizgi,twoside,paper,12pt]{article}

\newcommand{\ben}{\begin{enumerate}}
\newcommand{\een}{\end{enumerate}}
\newcommand{\be}{\begin{equation}}
\newcommand{\ee}{\end{equation}}
\newcommand{\bea}{\begin{eqnarray}}
\newcommand{\eea}{\end{eqnarray}}
\newcommand{\bc}{\begin{center}}
\newcommand{\ec}{\end{center}}
\newcommand{\vs}{\vspace}
\newcommand{\mb}{\mbox{\ }}

\begin{document}

\newpage
\thispagestyle{empty}
\bc
{\bf Integrability of
Kersten-Krasil'shchik coupled KdV-mKdV equations:
singularity analysis and Lax pair}
\ec
\mb \vs{1cm}
\bc
Ay\c{s}e Karasu(Kalkanl{\i})$^{*}$, Sergei Yu. Sakovich$^{**}$,\\
\'{I}smet Yurdu\c{s}en$^{*}$
\ec
\mb \vs{1cm}
\bc
{\em
$^{*}$Department of Physics,
      Middle East Technical University,\\
      06531 Ankara, Turkey \\
$^{**}$Institute of Physics,
       National Academy of Sciences,\\
       220072 Minsk, Belarus}
\ec
\vs{1cm}
\bc
{\bf Abstract}
\ec
The integrability of a coupled KdV-mKdV system is tested by
means of singularity analysis. The true Lax pair associated with
this system is obtained by the use of prolongation technique.
{\em
E-mail: akarasu@metu.edu.tr, saks@pisem.net,
ismety@newton.physics.metu.edu.tr}

\newpage

\setcounter{page}{1}
Very recently, Kersten and Krasil'shchik [1] constructed the
recursion operator for symmetries of a coupled KdV-mKdV system
\bea
u_t&=&-u_{xxx}+6uu_x-3ww_{xxx}-3w_x w_{xx}+3u_x w^2+6uww_x,\nonumber\\
w_t&=&-w_{xxx}+3w^2 w_x+3uw_x+3u_xw,
\eea
which arises as the classical part of one of superextensions of the
KdV equation. In this work, we study the integrability of this system
using the Painlev\'{e} test. Then, we use Dodd-Fordy [2] algorithm
of Wahlquist-Estabrook [3] prolongation technique in order to
obtain the Lax pair. We find a 3x3 matrix spectral problem
for the Kersten-Krasil'shchik system.\\ \\
{\em Singularity Analysis}:
Let us study the integrability of (1)
following the Weiss-Kruskal algorithm of singularity analysis
[4], [5]. The algorithm is well known and widely used, therefore
we omit unessential computational details.

First, we find that a hypersurface $\phi(x,t)=0$ is
non-characteristic for the system (1) if $\phi_x \neq 0$ and set
$\phi_x=1$ without loss of generality. Then we substitute the
expansions
\bea
u=u_0(t)\phi^{\alpha} + \ldots +
u_r(t)\phi^{r+\alpha} +\ldots, \nonumber\\
w=w_0(t)\phi^{\beta} + \ldots +
w_r(t)\phi^{r+\beta} +\ldots,
\eea
into (1), and find the
following branches (i.e. admissible choices of $\alpha, \beta, u_0$
and $w_0$), together with the positions $r$ of resonances (where
arbitrary functions can enter the expansions):
\bea
\alpha &=&-2,\; \; \beta=-1,\; \;u_0=1,\; \;w_0=\pm i, \nonumber\\
r &=& -1,1,2,3,4,6;
\eea
\bea
\alpha &=& -2,\;\; \beta=-1,\; \;u_0=2,\; \;w_0=\pm 2i, \nonumber\\
r &=& -2,-1,3,3,4,8;
\eea
\bea
\alpha &=& -2,\;\; \beta=2,\; \;u_0=2, \; \; \forall w_0(t), \nonumber\\
r &=& -4,-1,0,1,4,6;
\eea
\bea
\alpha &=& -2,\;\; \beta=3,\; \;u_0=2, \;\; \forall w_0(t), \nonumber\\
r &=& -5,-1,-1,0,4,6;
\eea
besides those which correspond to the Taylor expansions governed
by the Cauchy-Kovalevskaya theorem.

The branch (3) is generic: the expansions (2) with (3) describe
the behavior of a generic solution near its singularity.
The non-generic branches (4), (5) and (6) correspond to singularities
of special solutions. The branches (4) and (5) admit the following
interpretation, in the spirit of [6]: (4) describes the collision of two
generic poles (3) with same sign of $w_0$, whereas (5) describes the
collision of two generic poles (3) with opposite signs of $w_0$. The
branch (6) corresponds to (5) with $w_0 \rightarrow 0$.

Next, we find from (1) the recursion relations for the coefficients
$u_n(t)$ and $w_n(t)\; (n=0,1,2,\ldots)$ of the expansions (2),
separately for each of the branches, and check the consistency
of those recursion relations at the resonances. The recursion
relations turn out to be consistent, therefore the expansions
(2) of solutions of (1) are free from logarithmic terms. We
conclude that the system (1) passes the Painlev\'{e} test for
integrability successfully and must be expected to possess
a Lax pair.\\ \\
{\em Prolongation Structure}:
By introducing the variables $p = u_x$, $q = w_x$, $r = p_x$, $s = q_x$,
we assume that there exist NxN matrix functions  $F$ and $G$,
depending upon $u,w,p,q,r,s$, such that
\bea
y_x &=& - y F, \nonumber\\
y_t &=& - y G,
\eea
where $y$ is a row matrix with elements $y^A,
\; A =1,\ldots, N$.
The system of  equations in (1) can be represented as the
compatibility conditions of (7) if
\be
F_t - G_x + [F,G]=0,
\ee
where $[F,G]$ is the matrix commutator. This
requirement gives the set of partial differential equations
for $F$ and  $G$:
\bea
F_p =  F_q = F_r = F_s = 0 ,\qquad
F_u = -G_r  ,\qquad
3wF_u+F_w =- G_s , \nonumber \\
pG_u + qG_w +rG_p + sG_q - 3(2up-qs+pw^2+2uwq)F_u \nonumber\\
      -3(w^2q+uq+pw)F_w - [F,G] =0.
\eea
Next, we integrate equations (9) and find
\be
F=(uw-\frac{w^3}{2})X_1 +\frac{w^2}{2}X_2 + uX_3+ wX_4 + X_5,
\ee
where $X_1,X_2,X_3,X_4,X_5$ are constant matrices of integration.
It is immediately seen that $X_1$
is in the center of prolongation algebra [3]. Hence, we can take
it to be zero and find $G$ as,
\bea
G = (-r-ws-q^2+2u^2-w^4-w^2u)X_3 -(s-w^3-3uw)X_4 \nonumber\\
    -(p+wq)X_6 - uw X_7 -(\frac{w^2}{2}+u) X_8 \nonumber\\
    -q X_9 -\frac{w^2}{2} X_{10}- w X_{11} + X_0,
\eea
where $X_0$ is a constant matrix of integration.
The remaining elements are
\bea
X_6  =  [X_5,X_3],\qquad
X_7  =  [X_4,X_6], \qquad
X_8  =  [X_5,X_6], \nonumber\\
X_9  =  [X_5,X_4], \qquad
X_{10}  =  [X_4,X_9], \qquad
X_{11}  =  [X_5,X_9].
\eea
The integrability conditions impose the following restrictions on
$X_i$,\\$(i =0,  \ldots , 11)$,
\bea
[X_2,X_3] = 0 , \quad
[X_5,X_0] = 0 , \quad
[X_3,[X_3,X_6]] = 0 , \quad
[X_2,[X_4,X_3]] = 0 , \nonumber\\
{}[X_3,[X_4,X_3]] = 0, \quad
[X_3,[X_4,[X_4,X_3]]]=0,\quad
[[X_4,[X_4,X_3]],[X_3,X_6]]=0,\nonumber\\
{}2X_6 +[X_5,X_2]=0,\quad
[X_3,X_0]-[X_5,X_8]=0,\quad
[X_4,X_2] + 4 [X_4,X_3] = 0 , \nonumber\\
{}[X_4,X_0]-[X_5,X_{11}] = 0 , \quad
 3X_6 - \frac{1}{2}[X_5,[X_3,X_6]]-[X_3,X_8] = 0,\nonumber\\
{}3X_2 - 3[X_4,[X_4,X_3]]-[X_2,X_6] + [X_3,X_6] = 0,\nonumber\\
{}X_7 + 2[X_5,[X_4,X_3]]-[X_3,X_9]=0, \nonumber\\
{}[X_2,X_0]-2[X_4,X_{11}]-[X_5,X_8]-[X_5,X_{10}] = 0,\nonumber\\
{}[X_2,[X_5,[X_4,X_3]]]+[X_2,X_7]+
  \frac{1}{2}[X_2,[X_2,X_9]]=0,\nonumber\\
{}3X_9-[X_3,X_{11}]-[X_4,X_8]-[X_5,X_7]-
  2[X_5,[X_5,[X_4,X_3]]]=0,\nonumber\\
{}[X_3,X_7]+\frac{1}{2}[X_4,[X_3,X_6]]+
  [X_3,[X_5,[X_4,X_3]]]=0,\nonumber\\ \nonumber\\
 X_9-\frac{1}{2}([X_2,X_{11}]+[X_4,X_8]+
  [X_4,X_{10}])-\nonumber\\
 \frac{1}{3}([X_5,[X_5,[X_4,X_3]]]+
  [X_5,X_7])-\frac{1}{6}[X_5,[X_2,X_9] ]=0,\nonumber\\ \nonumber\\
 \frac{1}{2}[X_2,X_5]+\frac{1}{4}([X_2,X_8]+
  [X_2,X_{10}])+\nonumber\\
 \frac{1}{3}([X_4,X_7]+[X_4,[X_5,[X_4,X_3]]])+
  \frac{1}{6}[X_4,[X_2,X_9]]=0,\nonumber \\ \nonumber\\
 3X_6-\frac{1}{2}([X_2,X_8]+[X_3,X_8]+[X_3,X_{10}])-
  [X_4,X_7]-\nonumber\\
 2[X_5,[X_3,X_6]]-[X_4,[X_5,[X_4,X_3]]]-
  2[X_5,[X_4,[,X_4,X_3]]]=0,\nonumber \\ \nonumber\\
 8[X_4,X_3]+\frac{1}{4}[X_2,[X_2,X_9]]-
  2[X_4,[X_4,[X_4,X_3]]]-\nonumber \\
 \frac{1}{6}([X_3,[X_2,X_9]]+11[X_4,[X_3,X_6]])=0.\nonumber\\
\eea
Together with  the Jacobi identities we obtain further relations:
\bea
[X_2,X_6] +2 [X_3,X_6] = 0, \qquad
[X_4,X_{11}] - [X_5,X_{10}] = 0, \nonumber\\
{}[X_5,[X_3,X_6]] - [X_3,X_8] = 0,\qquad
[X_2,X_8] - [X_5,[X_2,X_6]]=0, \nonumber\\
{}[X_5,[X_4,X_3]]+[X_3,X_9] - X_7 = 0, \nonumber\\
{}-4[X_5,[X_4,X_3]] + [X_2,X_9] + 2X_7=0,\nonumber\\
{}[X_2,[X_5,[X_4,X_3]]]+2[[X_4,X_3],X_6]=0,\nonumber\\
{}[X_3,[X_5,[X_4,X_3]]]-[[X_4,X_3],X_6]=0,\nonumber\\
{}[X_3,[X_2,X_9]]-[X_2,[X_3,X_9]]=0,\nonumber\\
{}[X_4,X_3]=0, \qquad
  [X_2,X_7]=0, \qquad
  [X_3,X_7]=0, \nonumber\\
{}[X_3,X_{10}]=0, \qquad
  [X_4,X_7]=0, \qquad
  [X_5,X_7]=X_9,\nonumber\\
{}[X_2,[X_2,X_9]]=0,\qquad
  [X_4,[X_3,X_6]]=0,\nonumber\\
{}[X_5,X_8]+[X_5,X_{10}]=0.
\eea
In order to find the Lie algebra generated by F and matrix
representations of the generators $\{X_i\}_0^{11}$, we follow
the strategy of Dodd-Fordy [3].
First we reduce the number of elements. By using equations (12)
-(14), we get  $X_2 =-2X_3$. Next, we locate nilpotent and neutral
elements. The equations (12) and (13) together with  $X_2 =-2X_3$
give that, $[X_5,X_3]=X_6$ and
$[X_3,X_6]=2X_3$, hence $X_3$ is nilpotent and $X_6$ is the
neutral element.
Let us note that the system of equations in (1) has the following
scale symmetry
\be
x \rightarrow \lambda^{-1} x , \qquad
t \rightarrow \lambda^{-3} t , \qquad
u \rightarrow \lambda^{2} u ,\qquad
w \rightarrow \lambda w , \nonumber
\ee
which implies that the elements $X_{i}$ must satisfy
\bea
X_0 \rightarrow \lambda^{3}X_0 , \qquad
X_3 \rightarrow \lambda^{-1} X_3 , \qquad
X_4 \rightarrow  X_4 ,\qquad
X_5 \rightarrow \lambda X_5 ,\nonumber\\
{}X_6 \rightarrow X_6, \qquad
X_7 \rightarrow  X_7 ,\qquad
X_8 \rightarrow \lambda X_8,\qquad
X_9 \rightarrow \lambda X_9,\nonumber\\
{}X_{10} \rightarrow \lambda X_{10},\qquad
X_{11} \rightarrow \lambda^2 X_{11},
\eea
where $\lambda$ is a constant.
By using the basis elements,
we try to embed the prolongation algebra into  $sl(n+1,c)$. Starting
from the case $n=1$, we found that $sl(2,c)$  can not be
the whole algebra. The simplest non-trivial closure is in terms of
$sl(3,c)$. We take
\be
X_3 = e_{-\alpha_1}, \qquad
X_6 = h_1,
\ee
where we use the standart Cartan-Weyl basis [7] of $A_2$.
Together with the scale symmetries we find that
\bea
X_0 & = &-4c_2^2\lambda^4 e_{-\alpha_1}-36 c_1^3\lambda^3(h_1 + 2h_2)
     -4c_2\lambda^2e_{\alpha_1},\nonumber\\
X_4 & = & d_1(h_1+2 h_2)+
      d_2 \lambda^{-1}e_{\alpha_2}+
      d_3\lambda^2 e_{-\alpha_1-\alpha_2},\nonumber\\
X_5 & = & e_{\alpha_1}+ c_1\lambda(h_1 + 2h_2)
      +c_2\lambda^2e_{-\alpha_1},\nonumber\\
X_7 & = & d_2\lambda^{-1}e_{\alpha_2}+d_3\lambda^2
        e_{-\alpha_1-\alpha_2},\\
X_8 & = &-2e_{\alpha_1}+2c_2\lambda^2e_{-\alpha_1},\nonumber\\
X_9 & = & d_2\lambda^{-1}e_{\alpha_1+\alpha_2}-d_3\lambda^2 e_{-\alpha_2}
      +3c_1d_2 e_{\alpha_2}-3c_1d_3\lambda^3 e_{-\alpha_1-\alpha_2},
      \nonumber\\
X_{10} & = & -d_2 d_3\lambda(h_1+2 h_2)-
         6c_1d_2d_3\lambda^2 e_{-\alpha_1}, \nonumber\\
X_{11} & = & (9c_1^2+c_2)d_2\lambda e_{\alpha_2} +6c_1d_3\lambda^3
        e_{-\alpha_2}+ 6c_1d_2 e_{\alpha_1+\alpha_2}+
        (9c_1^2+c_2)d_3\lambda^4 e_{-\alpha_1-\alpha_2},\nonumber
\eea
where $\{c_i\}_1^2$ and $\{d_i\}_1^3$ are constants with conditions
\be
d_1d_2=0, \qquad
d_1d_3=0, \qquad
d_2d_3=6c_1,\qquad
c_2=9c_1^2.
\ee
We choose $d_1=0$, $c_1=d_2=1$.
So that, $X_7=X_4$ and $X_0=-36\lambda^2X_5$. Then, we obtain
the matrix representations of the generators $X_{i}$ as
\bea
X_3  = \left( \begin{array}{ccc}
0 & 0 & 0 \\
0 & 0 & 0 \\
1 & 0 & 0 \\
\end{array} \right),  \qquad
X_4  = \left( \begin{array}{ccc}
0 & 0 & 0 \\
-\lambda^{-1} & 0 & 0 \\
0 & 6\lambda^2 & 0
\end{array} \right) , \nonumber \\
{}X_5  = \left( \begin{array}{ccc}
-\lambda & 0 & 1 \\
0 & 2\lambda & 0 \\
9\lambda^2 & 0 & -\lambda \\
\end{array} \right) , \qquad
X_6  = \left( \begin{array}{ccc}
1 & 0 & 0 \\
0 & 0 & 0 \\
0 & 0 & -1 \\
\end{array} \right) , \nonumber \\
{}X_8  = \left( \begin{array}{ccc}
0 & 0 & -2 \\
0 & 0 &  0 \\
18\lambda^2 & 0 & 0 \\
\end{array} \right),  \qquad
X_9  = \left( \begin{array}{ccc}
0 & 6\lambda^2 & 0 \\
-3 & 0 & \lambda^{-1} \\
0 & -18\lambda^3 & 0  \\
\end{array} \right) , \nonumber \\
{}X_{10}  = \left( \begin{array}{ccc}
6\lambda & 0 & 0 \\
0 & -12\lambda & 0 \\
-36\lambda^2 & 0 & 6\lambda \\
\end{array} \right),\qquad
X_{11}  = \left( \begin{array}{ccc}
0 & -36\lambda^3 & 0  \\
-18\lambda & 0 & 6 \\
0 & 108\lambda^4 & 0 \\
\end{array} \right).
\eea
By substituting the matrix representations of the generators into
equations (10) and (11) we can construct the Lax pair,
$\Psi_x=X\Psi$,$\Psi_t=T\Psi$,
for the system (1), with the following matrices X and T:
\be
X=\left( \begin{array}{ccc}
\lambda & w \lambda^{-1} & w^2-u-9\lambda^2 \\
0 & -2\lambda & -6w\lambda^2 \\
-1 & 0 & \lambda \\
\end{array} \right),
\ee
$T=\{\{p+wq+3\lambda w^2-36\lambda^3,
     (w^3+2uw-s)\lambda^{-1}-3q-18\lambda w, r+ws+\\
     q^2-2u^2+w^4+w^2u-9\lambda^2 w^2+
     18\lambda^2u+324\lambda^4\},
     \{6q\lambda^2-36\lambda^3 w, -6\lambda w^2+\\72\lambda^3,
      6(s-w^3-2uw)\lambda^2-18q\lambda^3+108\lambda^4w\},
    \{-w^2-2u+36\lambda^2, q\lambda^{-1}+6w,
      -p-wq+3\lambda w^2-36\lambda^3\}\}$, where the matrix T
is written by rows and $X=-F^{\dagger}$,$T=-G^{\dagger}$,
$\Psi=y^{\dagger}$.

The forms of X and T are unusual in the sense of the dependence
on $\lambda$. It is possible to obtain equivalent matrices by
the gauge transformation,
\be
 X^{\prime}=SXS^{-1},\;\; T^{\prime}=STS^{-1},
\ee
where
\be
S=\left( \begin{array}{ccc}
1 & 0 & 0 \\
0 & 0 & -1 \\
0 & \lambda^{-1} & 0 \\
\end{array} \right).
\ee
The result is
\be
X^{\prime}=\left( \begin{array}{ccc}
\lambda & u-w^2+9\lambda^2 & w \\
1 & \lambda & 0 \\
0 & 6\lambda w & -2\lambda \\
\end{array} \right),
\ee
$T^{\prime}=\{\{p+wq+3\lambda w^2-36\lambda^3,
     -r-ws-q^2+2u^2-w^4-w^2u+9 \lambda^2 w^2-
      18\lambda^2u-324\lambda^4,
      w^3+2uw-s-3q\lambda-18\lambda^2 w\},
     \{w^2+2u-36\lambda^2,-p-wq+3\lambda w^2-36\lambda^3,
     -q-6w\lambda\},\{6q\lambda-36\lambda^2 w,
     -6(s-w^3-2uw)\lambda+18q\lambda^2-108\lambda^3 w,
     -6\lambda w^2+72\lambda^3\}\}$.\\
The matrix $X^{\prime}$ gives us exactly the spectral problem
for the KdV equation when $w=0$. But $X^{\prime}$
does not reduce to the one for mKdV equation when $u=0$.
This result should be expected because the
Kersten-Krasil'shchik system, when $u=0$, gives not only mKdV
equation, as stated in [1], but also an ordinary differential
equation in $w$. Finally, we note that the Lax pair obtained
from (7) with (24) is a true Lax pair since the parameter
$\lambda$ cannot be removed from $X^{\prime}$ by a gauge
transformation, as can be proven by a gauge-invariant
technique [8].\\
\baslik{Acknowledgements}
This work is supported in part by
the Scientific and Technical Research Council of Turkey (TUBITAK).

\begin{reference}
\item P. Kersten, J. Krasil'shchik, E-print(2000) arXiv:nlin.SI/0010041.
\item R. Dodd, A. Fordy, Proc.R.Soc.Lond., {\bf A 385},
 (1983)389.
\item H.D. Wahlquist, F.B. Estabrook,
 J.Math.Phys.,{\bf 16}, (1975)1.
\item J. Weiss, M. Tabor, G. Carnevale,
 J.Math.Phys.,{\bf 24}, (1983)522.
\item M. Jimbo, M.D. Kruskal, T. Miwa,
 Phys. Lett. {\bf A 92}, (1982)59.
\item A.C. Newell, M. Tabor, Y.B. Zeng,
 Physica {\bf D 29},(1987)1.
\item J.E. Humphreys, {\bf "Introduction to Lie algebras
 and representation theory"}, Springer-Verlag, NewYork, 1972.
\item S. Yu. Sakovich, J. Phys. {\bf A 28}, (1995)2861.

\end{reference}

\end{document}